\documentstyle[sprocl]{article}

\input{psfig}
\def\Journal#1#2#3#4{{#1} {\bf #2}, #3 (#4)}

\def\NPB{{\em Nucl. Phys.} B}

\def\PRD{{\em Phys. Rev.} D}
\def\ZPC{{\em Z. Phys.} C}
\def\be{\begin{equation}}
\def\ee{\end{equation}}
\def\bea{\begin{eqnarray}}
\def\eea{\end{eqnarray}}
\begin{document}

\rightline{{\bf IEM--FT--144/96}}
\rightline{{\bf SUSX--TH--96--015}}
\vskip0.8cm

\title{RELIC ABUNDANCES AND DETECTION RATES OF NEUTRALINOS
IN STRING-INSPIRED SUPERGRAVITY MODELS\footnote{To appear in the
Proceedings of the International Conference on the Identification of Dark
Matter, {\em IDM'96}, Sheffield (UK), 8--12 September 1996.}}

\author{ B. de CARLOS }

\address{Centre for Theoretical Physics,\\
University of Sussex,\\
Falmer, Brighton BN1 9QH, UK}

\author{ G.V. KRANIOTIS}

\address{Royal Holloway and Bedford New College,\\
University of London,\\
Egham, Surrey TW20 0EX, UK}


\maketitle
\abstracts{
We calculate {\it relic abundances} and {\it detection rates} of the 
neutralino (LSP) in string-inspired supergravity models
with {\it dilaton-moduli} induced supersymmetry breaking.
In particular we investigate {\it universal} scenarios for
the soft-supersymmetry breaking terms from Calabi-Yau compactifications,
as well as from the dilaton-dominated limit. {\it Non-universal}
scenarios from orbifold string theory are also incorporated into
the analysis. In all cases, in the cosmologically interesting region,
we find $m_{LSP}\geq 50$ GeV and direct-detection rates in the range 
O($10^{-3}$ events/(Kg day))--O($10^{-4}$ events/(Kg day)).
{\it Indirect-detection} rates from LSPs captured in the Sun
are also calculated.}

String theory is the leading candidate for the unification
of the four fundamental interactions, gravitation plus
gauge forces. Supersymmetry (SUSY), which is a {\it hot} experimental
target of current accelerators and the future Large Hadron Collider
(LHC), is naturally embedded in string theory.  
Besides solving in the technical sense the gauge hierarchy
problem, SUSY provides us with a significant bonus: the
LSP in string-inspired supergravity models with
$R$-symmetry conservation is {\it stable} and it is
an ideal candidate for dark matter. 
The detection of such a particle will 
constitute an overwhelming
evidence for SUSY and {\it non-baryonic} dark matter.
Since a lot of progress has been achieved in experiments
designed to detect the LSP, and new strategies for ongoing
and planned experiments have been decided,
the study of relic abundances and detection rates of LSPs in 
string-inspired SUSY models
is very well motivated and constitutes our contribution to this
conference.

However, a satisfactory SUSY breaking mechanism in string 
theory is still lacking. As a result, a definite prediction
for SUSY masses such as $m_{LSP}$ is not possible yet.
The introduction of soft-susy breaking terms
such as gaugino masses $M_a$, scalar masses $m_{\alpha}$,
trilinear scalar soft terms $A_{\alpha\beta\gamma}$ and
bilinear soft terms $B_a$ in order to make SUSY theories realistic,  
discussed in this workshop by J.Ellis and
A.Bottino\cite{ELLBO}, parametrizes
our ignorance of the exact mechanism of SUSY breaking.
Recently progress has been reported in deriving soft-terms
from string theory \cite{Iba:Spain}.
In this work the effect of SUSY-breaking is parametrized by the
vacuum expectation values ({\it vevs}) of the $F$-terms associated to 
the dilaton ($S$) and the moduli 
($T_{m}$) chiral superfields, generically
present in large classes of four-dimensional supersymmetric heterotic
strings. This attempt is an important step towards a theory of
soft terms and for the extraction of model independent phenomenology
from string theory. 
In this framework, the stringy soft-terms,
in {\it no-scale} scenarios with
zero cosmological constant depend on the
gravitino mass $m_{3/2}$, and the Goldstino angle $\theta$
which specifies the extent to which the source
of supersymmetry breaking resides in the dilaton
versus moduli sector.
This gives rise to various scenarios for the soft terms at the string 
scale $M_{str}$, among which we will consider here the 
following\footnote{For an analysis of the phenomenology of dark 
matter in other superstring inspired scenarios see \cite{chen}.}: 

\begin{itemize}
\item In the large $T-$ limit of {\it Calabi-Yau} compactifications
we have
\begin{eqnarray}
m_{\alpha}^{2}&=&m^{2}_{3/2}\sin^2\theta  \nonumber \\
M_a&=&\sqrt{3}\frac{k_a {\rm Re} S}{{\rm Re} f_a} m_{3/2}\sin\theta 
\nonumber \\
A_{\alpha\beta\gamma}&=&-\sqrt{3}m_{3/2}\sin\theta \label{calb} \\
B_{\mu}&=&m_{3/2}[-1-\sqrt{3}\sin\theta-\cos\theta]
=A-m_{3/2}(1+\cos\theta) \;\;, \nonumber
\end{eqnarray}
where $k_a$ is the Kac--Moody level associated to the corresponding
gauge group, and $f_a$ is the gauge kinetic function. Note that the
soft terms are, in this case, {\it universal} (i.e. {\it all} scalar
masses and {\it all} gaugino masses are the same at $M_{str}$).

\item 
If the dilaton $F$-term dominates in the process of SUSY breaking, i.e
$\sin\theta=1$, we get another {\it universal} scenario for the 
soft terms, the so--called {\it Dilaton Dominated}:
\begin{equation}
M_a=M_{1/2}=-A,\;\;\;m_0=\frac{1}{\sqrt{3}}M_{1/2}
\label{dil}
\end{equation}

\item However also {\it non-universal} scenarios arise in some of the 
models. For instance in the
O-I orbifold scenario the relevant soft-terms 
are given by
\begin{eqnarray}
m_{\alpha}^2&=&m^2_{3/2}(1+n_{\alpha} \cos^2\theta) \nonumber \\
A_{\alpha\beta\gamma}&=&-\sqrt{3}m_{3/2}\sin\theta-m_{3/2}
\cos\theta(3+n_{\alpha}+n_{\beta}+n_{\gamma}) \nonumber \\
M_a&=&\sqrt{3}m_{3/2} \frac{k_a {\rm Re}S}{{\rm Re} f_a} \sin\theta+
m_{3/2}\cos\theta \frac{B_a^{'}(T+T^{\ast})\hat{G}_{2}(T,T^{\ast})}
{32{\pi}^{3}{\rm Re}{f}_{a}} \nonumber \\
B_{\mu}&=&m_{3/2}[-1-\sqrt{3}\sin\theta-\cos\theta(3+n_H+n_{\hat{H}})] \;\;,
\label{vathmos}
\end{eqnarray}
where in Eq.~(\ref{vathmos}) the quantities that parametrize the lack
of universality for scalars, $n_{\alpha}$, are the modular weights
of the different fields (i.e. their charges under the $T-duality$ 
string symmetry); their numerical values (which are usually
negative integers) together with Re~$T$ are chosen so that the 
interactions are unified at $M_{U}\approx 2 \times 10^{16}$ GeV. All other 
quantities appearing in (\ref{vathmos}) can be found
in \cite{Iba:Spain}.
 
\end{itemize}

Using the above scenarios Eqs.~(\ref{calb})-(\ref{vathmos}) for the 
soft terms as boundary conditions at $M_{str}$, 
%for the relevant renormalization group equations,
one can calculate the physical spectrum at the weak scale by solving 
the renormalization
group equations for the masses of the different SUSY particles subject to
combined constraints coming both from the experiment and from imposing
a correct radiative electroweak symmetry breaking. The latter is 
enforced by minimizing the full one-loop effective potential.
The LSP and dark matter candidate in these models
neutralino (${\chi}_1^0$), which is a linear combination of the
superpartners of the neutral electroweak gauge bosons and of the two 
neutral Higgs fields:
\begin{equation}
\chi_1^0 = a_1 \tilde{B} + a_2 \tilde{W}^3 + a_3 \tilde{H}_1^0 + a_4 
\tilde{H}_2^0
\end{equation}
The relic abundance of the LSP, $\Omega_{\chi}h^2$,
is proportional to the thermally averaged cross 
section $<\sigma_{ann}v>$,
\begin{equation}
\Omega_{\chi}h^2 \propto \frac{1}{<\sigma_{ann}v>}
\label{reli}
\end{equation}
\begin{figure}
\centerline{
\psfig{figure=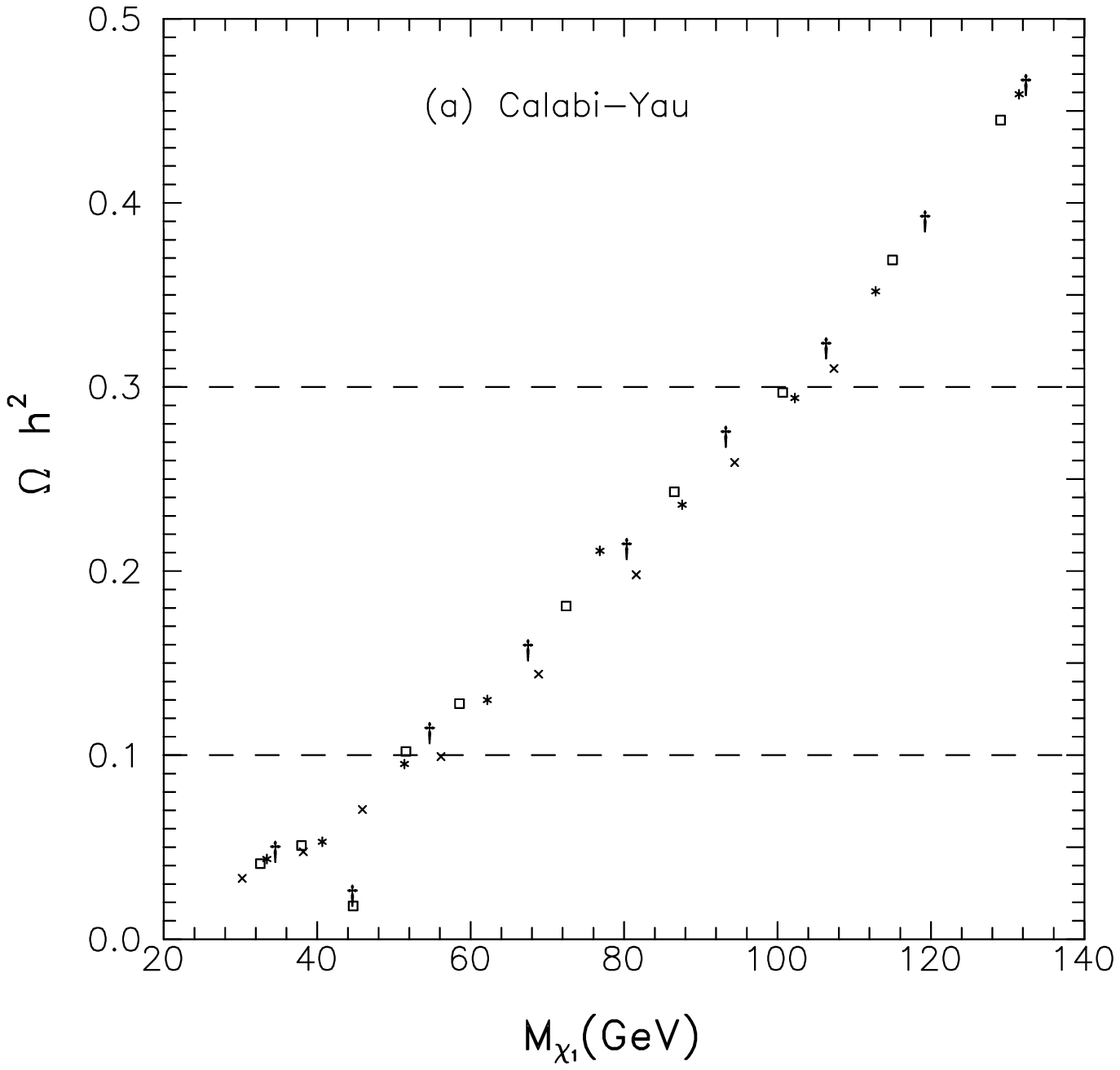,height=5.5cm,bbllx=3.cm,bblly=3.5cm,bburx=18cm,bbury=16cm}\
\psfig{figure=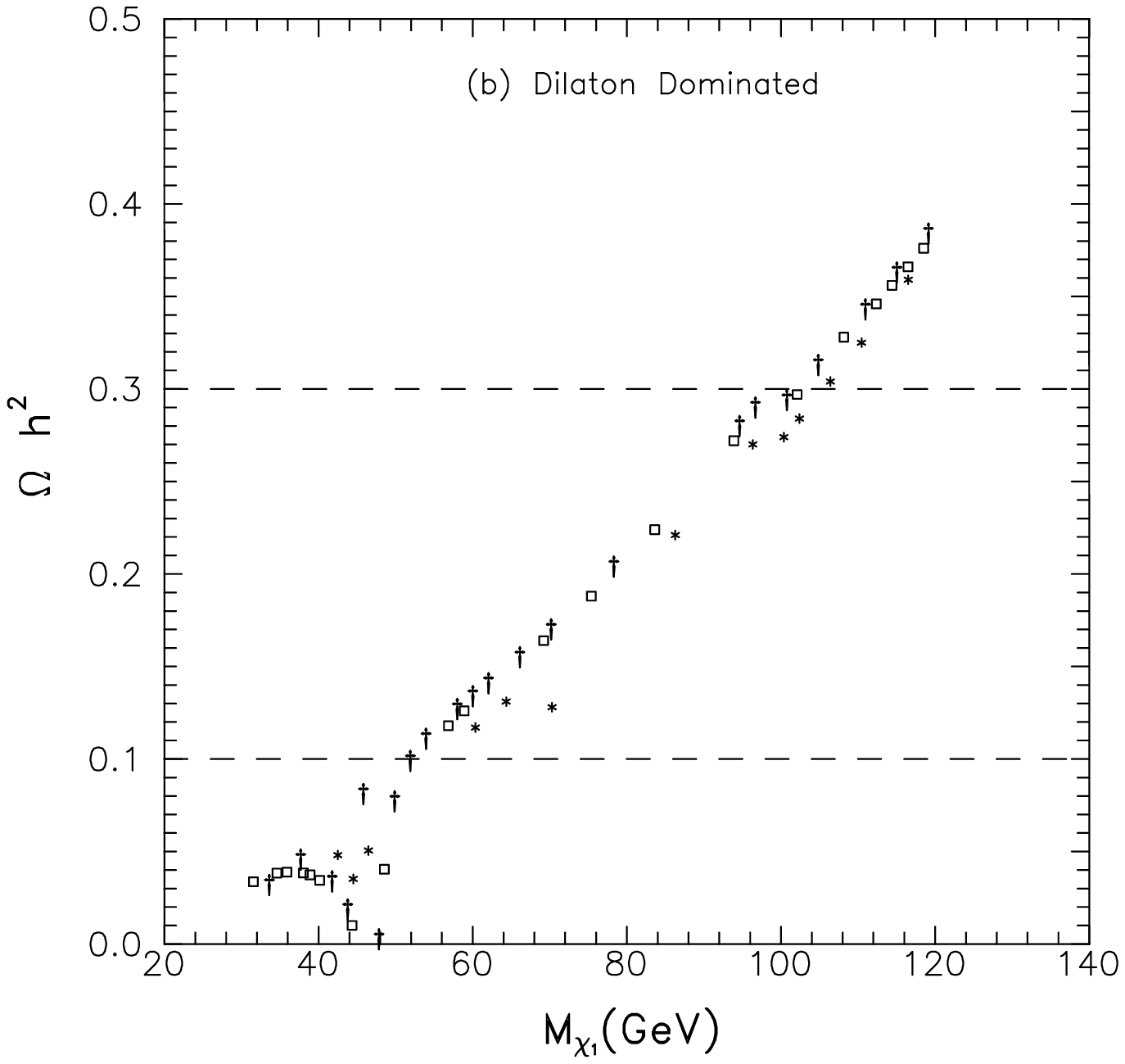,height=5.5cm,bbllx=3cm,bblly=3.5cm,bburx=18cm,bbury=16cm}
}
\caption{Plot of the relic abundance of the LSP (i.e. lightest
neutralino) vs its mass (in GeV) for (a) large T--limit of Calabi-Yau
compactifications (``x'' corresponds to $\theta=\pi/4$, ``$\ast$'' to 
$\theta=\pi/2$, ``$\dagger$'' to $\theta=3\pi/4$ and ``$\Box$'' to 
$\theta=7\pi/8$); (b) Dilaton Dominated models (where ``$\ast$''
corresponds to $\tan\beta=2.5$, ``$\dagger$'' to $\tan\beta=6$ and
``$\Box$'' to $\tan\beta=10$). The limits on  
$\Omega_{\chi}h^2$ are represented by the dashed lines.
}
\end{figure}
\begin{figure}
\centerline{
\psfig{figure=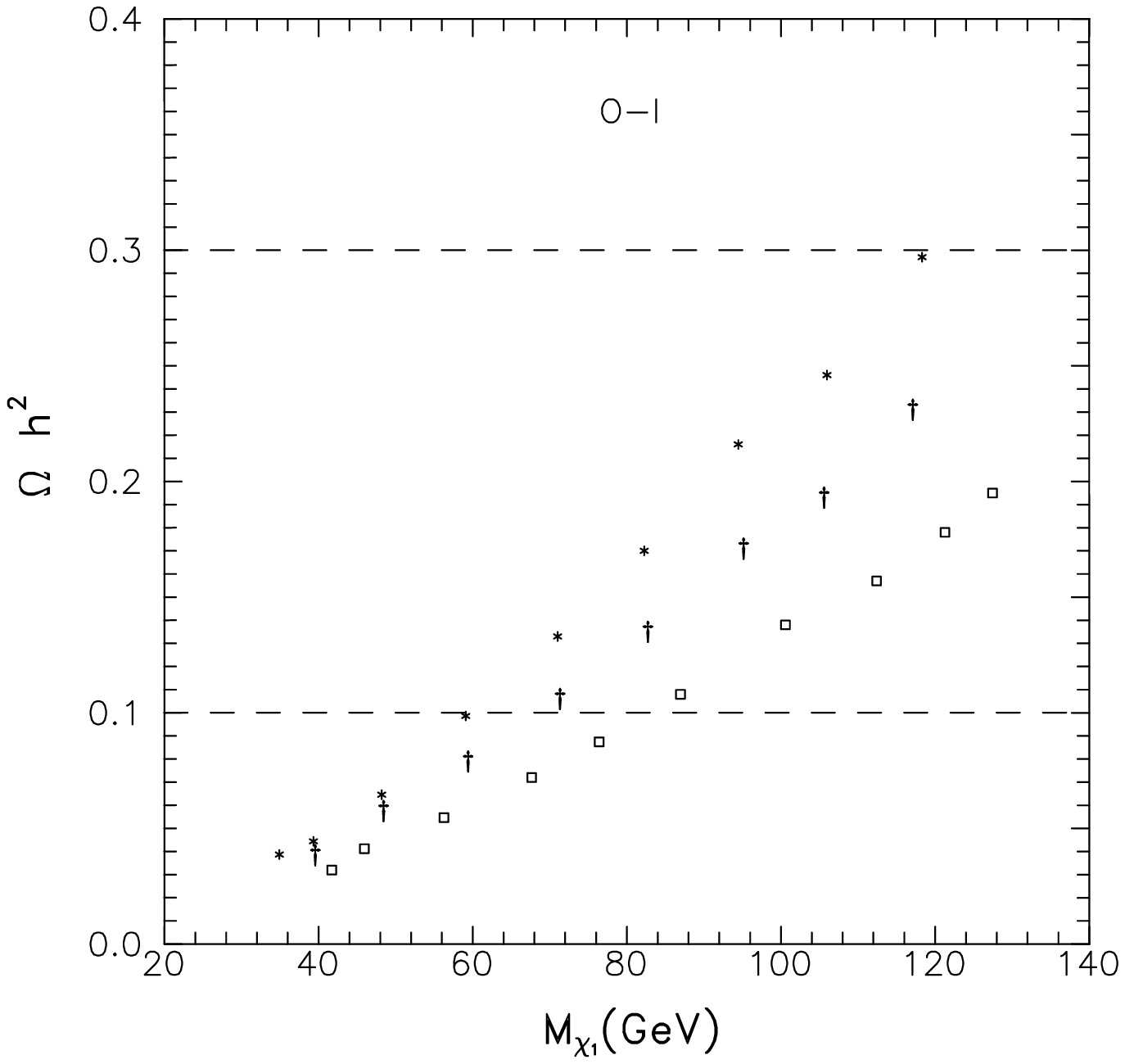,height=5.5cm,bbllx=3.cm,bblly=3.5cm,bburx=18cm,bbury=16cm}\
\psfig{figure=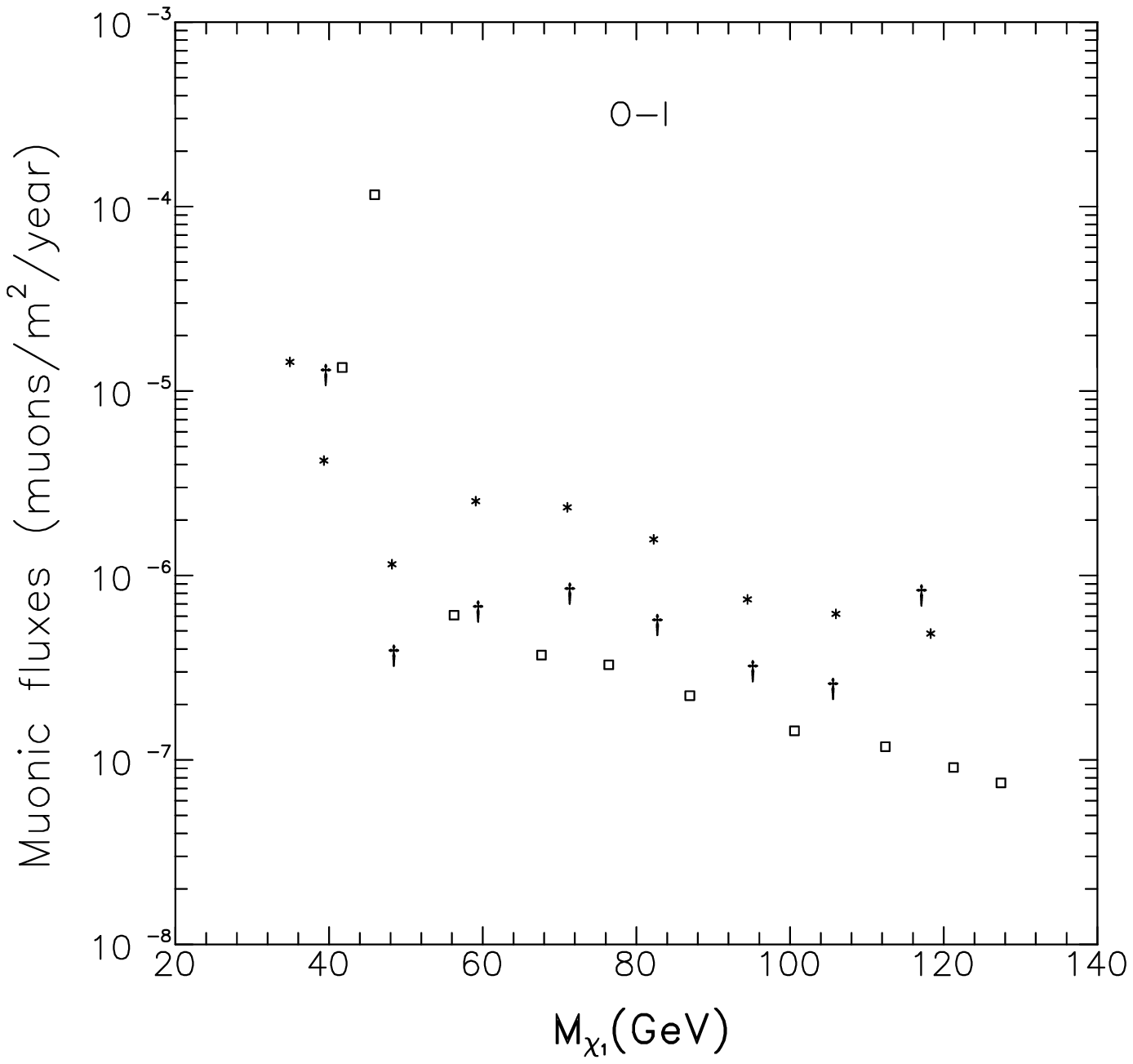,height=5.5cm,bbllx=3cm,bblly=3.5cm,bburx=18cm,bbury=16cm}
}
\caption{Plots of the relic abundance of the LSP (i.e. lightest
neutralino) (left) and the indirect detection rates of muon neutrinos
emerging from the Sun (right) vs its mass (in GeV) for the O-I model 
(here ``$\ast$'' corresponds to $\theta=\pi/2$, ``$\dagger$'' to 
$\theta=5\pi/8$ and ``$\Box$'' to $\theta=11\pi/16$). The limits on  
$\Omega_{\chi}h^2$ are represented by the dashed lines.
}
\end{figure}

In the evaluation of $<\sigma_{ann}v>$ we have considered
the following set of final states: fermion-antifermion
pairs, pairs of gauge bosons, Higgs-gauge boson pairs, pairs
of Higgs bosons and also $s$-wave contributions
for the two-gluon ($gg$) and ($q\bar{q}g$)
final states in the spirit of \cite{Jungman}
\footnote{We thank M. Kamionkowski
for providing us with the source code of Neutdriver.}. We enforce
the following constraints on the neutralino relic density\footnote{For
a more detailed discussion about constarints on $\Omega_{\chi} h^2$
see \cite{us}.} \cite{const}:
\begin{equation}
0.1 \leq \Omega_{\chi} h^2 \leq 0.3
\label{cosmo}
\end{equation}
As we can see from Figs.~1 (universal models) and 2 (non-universal
ones), the lower bound on $\Omega_{\chi} h^2$ implies a lower limit 
on $M_{\chi_1^0}$ of order $50$ GeV. Further reductions on the 
phenomenologically viable parameter space can be obtained by imposing 
FCNC constraints on these spectra, such as that they give rise to a 
branching ratio for the process $b \rightarrow s, \gamma$ within its
experimental limits \cite{us,drees,george}.

\begin{figure}
\centerline{
\psfig{figure=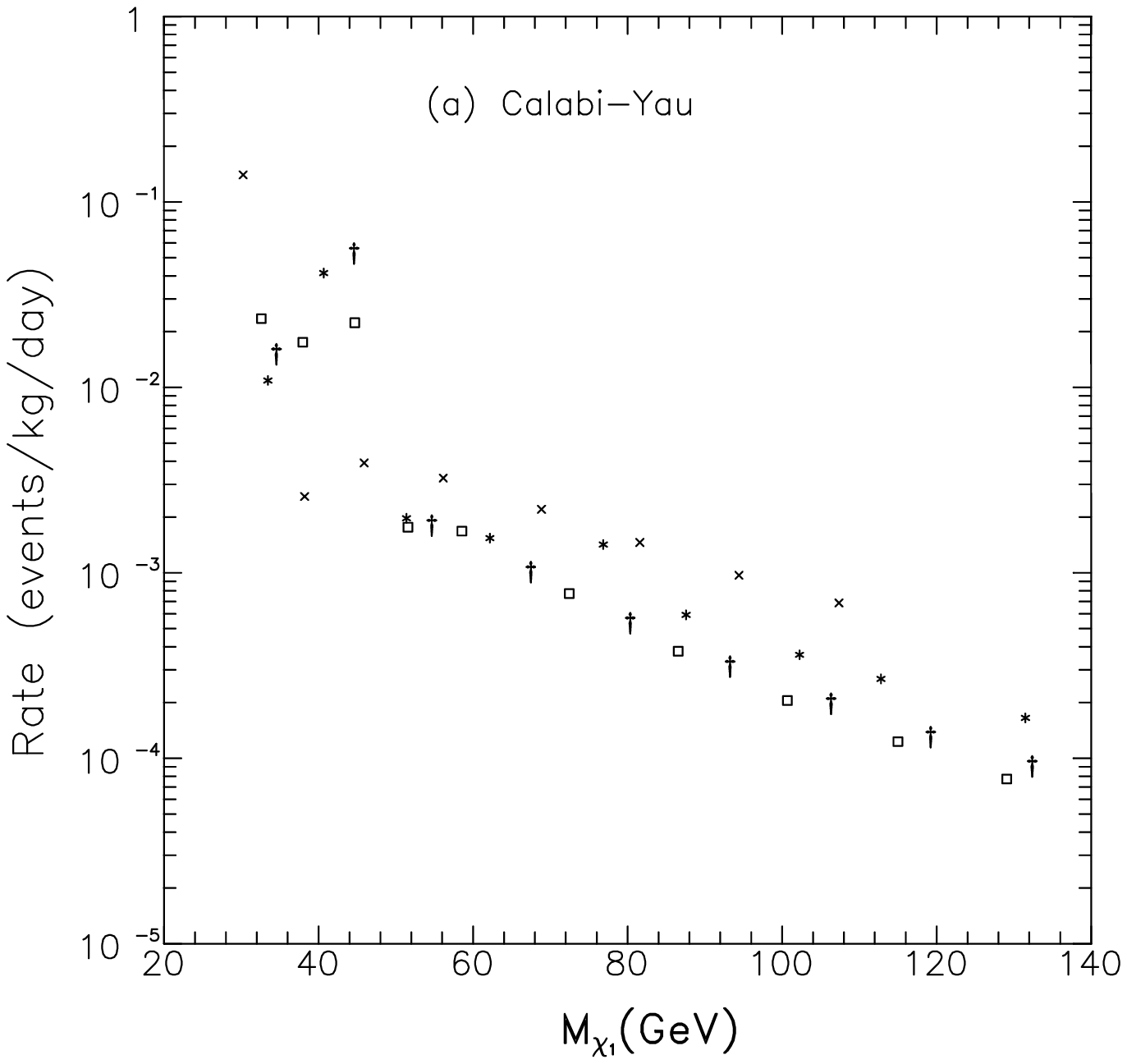,height=5.5cm,bbllx=3.cm,bblly=3.5cm,bburx=18cm,bbury=16cm}\
\psfig{figure=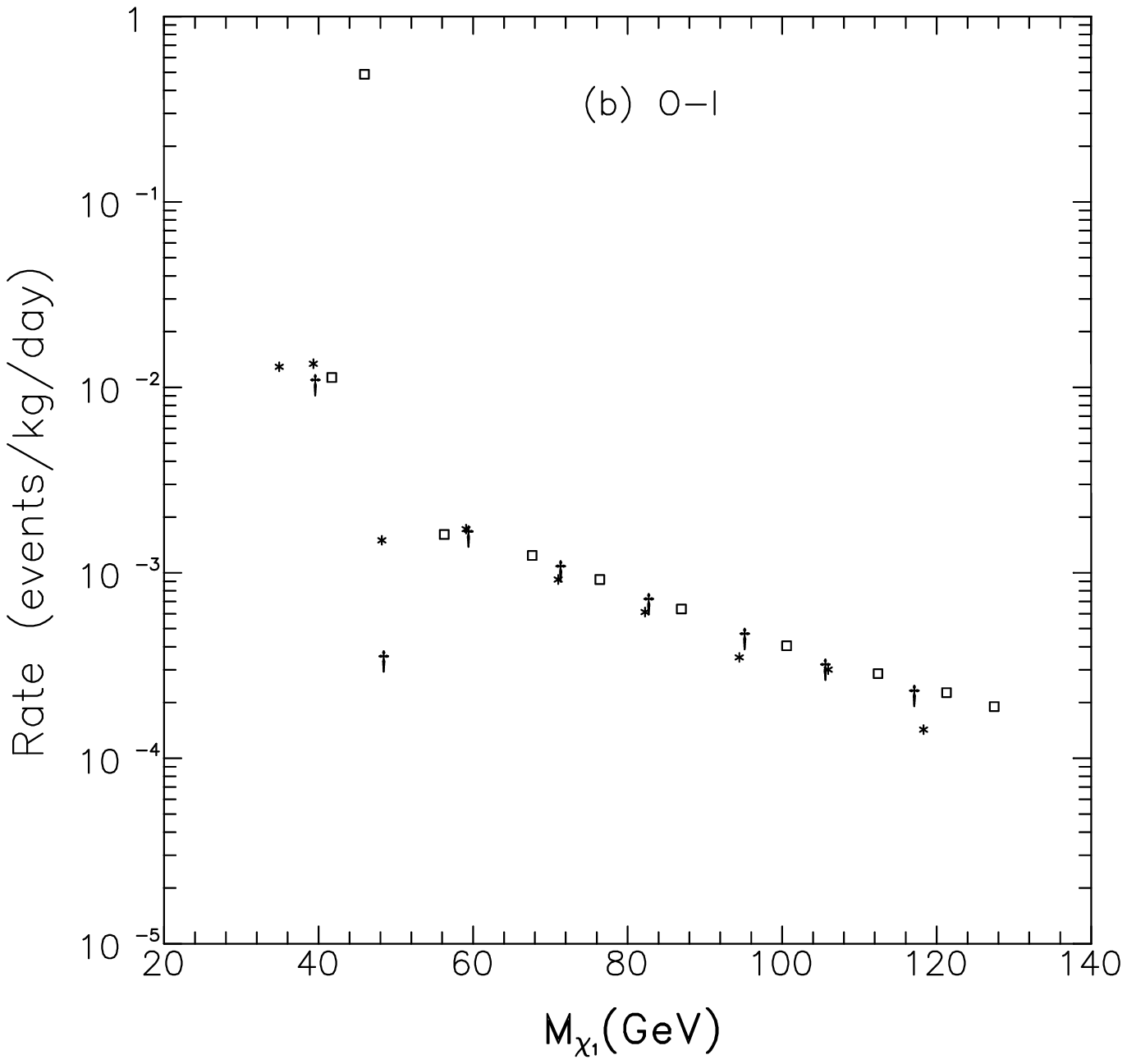,height=5.5cm,bbllx=3cm,bblly=3.5cm,bburx=18cm,bbury=16cm}
}
\caption{Plot of the direct detection rates of the LSP (i.e. lightest
neutralino) vs its mass (in GeV) for (a) large T--limit of Calabi-Yau
compactifications; (b) O-I model. The labels are the
same as in Figs.~1,2.
}
\end{figure}
Let's turn now to the detection rates of the LSP in these
models. There are two ways to detect the neutralino:
{\it direct detection} experiments try to observe a nucleus recoil
after an LSP-nucleus scattering;
in this case the detection rate is proportional
to the {\it local} LSP density 
$\rho_{\chi}$ as well as the elastic
cross section $\sigma_{elastic}$ of the LSP with a given
nucleus. 
$\sigma_{elastic}$
has two contributions: a coherent contribution,
due to Higgs and squark $\tilde{q}$
exchange diagrams, which depends
on $A^{2}$, $A$ being the mass number of the nucleus;
a spin-dependent contribution, arising from $Z$ 
and $\tilde{q}$ exchange again, proportional to the
total angular momentum $\lambda^2 J(J+1)$. 
The differential detection rate is given by
\begin{equation}
\frac{dR}{dQ}=\frac{\sigma_{elastic}\rho_{\chi}}
{4v_e m_{LSP}m^2_r} F^2(Q)[erf(\frac{v_{min}+v_e}{v_0})-
erf(\frac{v_{min}-v_e}{v_0})]
\label{diff}
\end{equation}
where all the relevant quantities are defined in \cite{Jungman}.
In Fig.~3 we present results for the
integrated rate $R$, i.e the number of events per
kilogram of $^{76}$Ge detector material per day. It can be seen that, 
in the cosmologically interesting region ($0.1< \Omega_{\chi} h^2<0.3$), the 
detection rates are in the range of 
O($10^{-3}$ events/(Kg day))--O($10^{-4}$ events/(Kg day)). 
Note that the highest detection rates are obtained 
in the region of
very small relic densities, i.e $\Omega_{\chi} h^2\leq 0.05$.

Indirect detection experiments try to measure the flux
of neutrino induced muons from captured LSPs in the Sun
or Earth. In Fig.~2 we present results for
muonic fluxes resulting from captured neutralinos in the 
Sun in the case of O-I model. In this case the resulting
rates are far below the current experimental sensitivity.

Therefore we can conclude that by
combining  cosmological constraints on $\Omega_{\chi} h^2$
with correct radiative electroweak symmetry breaking and experimental
limits on SUSY masses in string-ispired supergravity models,
one obtains a strong lower limit on the neutralino mass
i.e $M_{\chi}\geq 50$GeV.

Also given the optimism expressed by our experimental colleagues
in this workshop that
the desirable experimental sensitivity 
will be reached in shorter time 
we can say that WIMPs experiments will definetely
help in testing the validity of the assumption of
dilaton-moduli induced SUSY breaking in string theory.

\section*{Acknowledgments}
We thank Ed Copeland for encouraging us to present our results at this
workshop. GVK thanks the organizers for their warm hospitality and for
creating such an enjoyable atmosphere. 
He also thanks C.E. Vayonakis
for encouragement and many fruitful
discussions. The work of BdeC was supported 
by a PPARC Postdoctoral Fellowship.


\section*{References}
\begin{thebibliography}{99}


\bibitem{ELLBO} See J. Ellis, these same proceedings; \\
See A. Bottino, these same proceedings.

\bibitem{Iba:Spain} A. Brignole, L.E. Ib\'a\~nez, C. Mu\~noz, 
\Journal{\NPB}{422}{126}{1994}; erratum {\em ibid.}, B {\bf 436} 747 
(1995).

\bibitem{chen} C.-H. Chen, M. Drees, J.F. Gunion, hep-ph/9607421.

\bibitem{Jungman} G. Jungman, M. Kamionkowski, K. Griest, {\em Phys. Rep}
{\bf 267} 195 (1996).

\bibitem{us} B. de Carlos, G.V. Kraniotis, in preparation.

\bibitem{const} V. Berezinskii {\it et al}, {\em Astropart. Phys.} 
{\bf 5} 1 (1995); \\
V. Berezinskii {\it et al}, hep-ph/9603342.

\bibitem{drees} F.M. Borzumati, M. Drees, M.M. Nojiri, 
\Journal{\PRD}{51}{341}{1995}; \\
L. Bergstr\"om, P.Gondolo, hep-ph/9510252.

\bibitem{george} G.V. Kraniotis, \Journal{\ZPC}{71}{163}{1996}.
 
\bibitem{bd} C.D. Buchanan {\it et al}, \Journal{\PRD}{45}{4088}{1992}.

\end{thebibliography}
\end{document}